# Electrostatic Landscape of a H-Silicon Surface Probed by a Moveable Quantum Dot

*Taleana Huff[1,3], Thomas Dienel[1], Mohammad Rashidi[1], Roshan Achal[1,3], Lucian Livadaru[3], Jeremiah Croshaw[1], and Robert A. Wolkow[1,2,3]*

[1]Department of Physics, University of Alberta, Edmonton, Alberta, T6G 2J1, Canada

[2] Nanotechnology Research Centre, National Research Council Canada, Edmonton, Alberta, T6G 2M9, Canada

[3]Quantum Silicon, Inc., Edmonton, Alberta, T6G 2M9, Canada

**With nanoelectronics reaching the limit of atom-sized devices, it has become critical to examine how irregularities in the local environment can affect device functionality. Here, we characterize the influence of charged atomic species on the electrostatic potential of a semiconductor surface at the sub-nanometer scale. Using non-contact atomic force microscopy, two-dimensional maps of the contact potential difference are used to show the spatially varying electrostatic potential on the (100) surface of hydrogen-terminated highly-doped silicon. Three types of charged species, one on the surface and two within the bulk, are examined. An electric field sensitive spectroscopic signature of a single probe atom reports on nearby charged species. The identity of one of the near-surface species has been uncertain. That species, suspected of being boron or perhaps a negatively charged donor**

**species, we suggest is of a character more consistent with either a negatively charged interstitial hydrogen or a hydrogen vacancy complex.**

The ultimate miniaturization of technology will be constructed of individually placed atoms or molecules. Many notable studies have been presented including atomic-spin based logic,[1] molecular devices,[2–4] patterned dangling bond devices and structures,[5–8] single atom transistors,[9] probabilistic finite state machines,[10] and qubits.[11–13] Vital for all of these applications is a precise knowledge of the local electrostatic environment;[14] maintaining a regular electrostatic background on a scale comparable to the device or device component size is necessary to prevent aberrant behavior. Using the charge sensitivity of non-contact atomic-force microscopy (nc-AFM), we study local variations in the electrostatic environment at the surface of highly arsenic-doped hydrogen-terminated silicon (H-Si).

We apply two established AFM techniques. First, we use grid-based Kelvin probe force microscopy (KPFM)[15–17] maps to resolve electrostatic variations on a nm length scale. This technique has been previously used to examine charge distribution within single molecules[15,18] and in the proximity of surface species[19,20] by extracting the local contact potential difference for each point in an area, generating a 2D surface map of the potential energy landscape (See Methods-KPFM Maps). Here we apply this technique to the surface of a heavily doped semiconductor. We investigate two charged near-surface species of different character that display opposite electrostatic contrast.

Second, we apply a variant of scanning quantum dot microscopy[21,22] to individually probe the surrounding of the charged species. Scanning quantum dot microscopy typically relies on functionalizing the apex of the tip of an atomic force microscope with a quantum dot. Subtle changes of the quantum dot's charging behavior are employed to sense the electrostatic field emanating from the scanned surface.[21,22] Due to the functionalization of the tip with large clusters or organic molecules and the required large bias range applied during probing (up to several volts), this methodology has been classified as a "far-field" method with moderate spatial resolution.[22] Instead of functionalizing the tip's apex, here we "move" a single-atom quantum dot across the surface, enhancing the spatial resolution for our sample system.

We use dangling bonds (DBs) on the otherwise hydrogen-terminated silicon surface as our moveable quantum dots.[6,23,24] DBs possess gap states that are electronically isolated from the silicon bulk. DBs can hold zero, one, or two electrons resulting in a positive, neutral, or negative charge state respectively.[5,8,25,26] In correspondence with these three charge states, there exist two distinct charge transition levels, (+/0) and (0/-), the specific energies of which are sensitive to their local electrostatic environment.[6,24,25] Other articles have detailed the precise patterning [27–29] and erasing [30,31] of DBs on H-Si (see Methods), which we employ here to progressively march a probing DB through the vicinity of charged species (*e.g.* a second DB or a near-surface dopant atom). Moving the probe DB with respect to a charge of fixed position results in a bias shift at which the probe DB's (0/-) charge transition is measured by KPFM (*Δf(V))* spectroscopy, See Methods-KPFM Maps).[6,32–34] Examination of the observed shifts allows us to determine the sign and location of the fixed charge within the substrate, as well as the effective local dielectric constant and local screening length in the vicinity of the charge.

The identities of two species are firmly established in the literature as an ionized arsenic atom (a positive species) and a negatively charged DB. Previous work has suggested that a third may be either a boron contaminant atom or negatively charged arsenic atom.[35,36] We reconsider these assignments and find that the third species is more likely a negatively charged interstitial hydrogen atom or a hydrogen-vacancy complex.

**Concentration of Charged Features**

Scanning tunneling microscopy (STM) images of the same surface area of H-Si(100) 2×1 under three different tunneling conditions are displayed in Figure 1a-c. Figure 1a shows a typical filled states STM image with the characteristic rows of paired silicon atoms (dimers) running horizontally across the frame [37,38] wherein each surface silicon atom is capped by a single hydrogen atom. The scattered, irregularly shaped dark areas are etched pits (missing silicon atoms of the top layer), and the small bright protrusions are natively occurring DBs. The features highlighted by dashed circles (labeled T1-Type 1 and T2-Type 2) are the two near-surface charged species being examined in this work. While Type 1 features are known to be arsenic dopants, [35,39–41] the identity of Type 2 features remains less certain.[35,39,42] The appearance of the arsenic dopants strongly depends on their depth, the imaging orbital of the STM tip, and the lattice site the dopant occupies.[40] Similarly, variations in STM topography are reported for T2s, suggesting that they can reside in different near-surface layers,[35] but with less variety observed compared to arsenic dopants. The correlation between STM appearances and the charge states of the dopant atom will be discussed in detail later on.

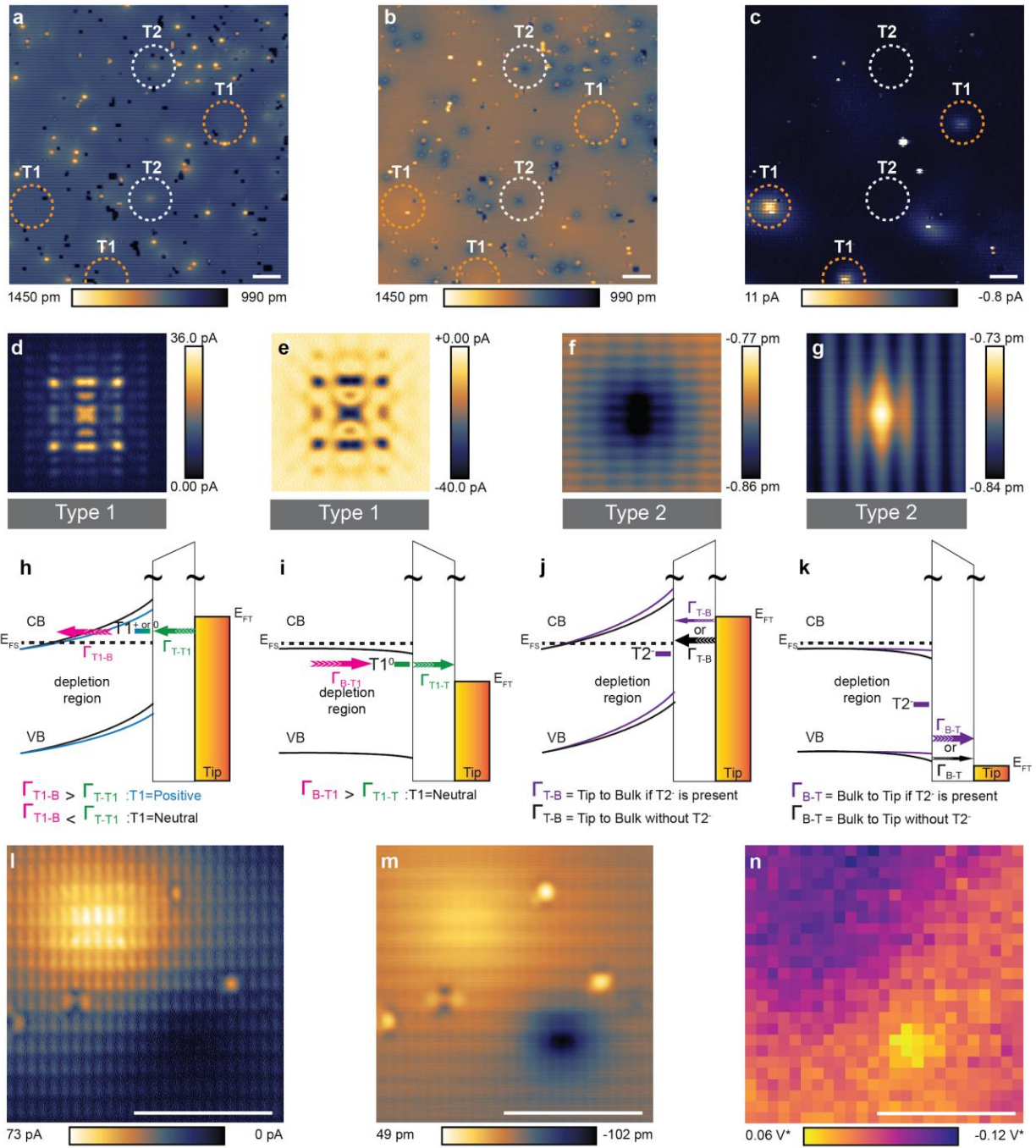

**Figure 1.** Identifying charged near-surface species. **(a)** Constant current filled states (V = -1.6 V and I = 50 pA) and **(b)** empty states (V = 1.3 V and I = 50 pA) STM images of the hydrogen-terminated silicon surface. **(c)** Constant height STM image ($z_{rel}$ = -150 pm and V = 300 mV) of the same area. The two different near-surface charged species are marked. All images are 50×50 nm$^2$ with $z_{rel}$ referenced to a common STM set-point defined by -1.8 V and 50 pA over a hydrogen-terminated silicon atom. **(d)** The arsenic in constant current mode at moderate positive bias ($z_{rel}$ = -300 pm and V = 300 mV). **(e)** The arsenic imaged in constant height STM mode with modest negative bias ($z_{rel}$ = -300 pm, V = -0.9 V). The T2 remains negative for all its constant current filled **(f)** and empty **(g)** states STM images. Images (d-g) are all 5×5 nm$^2$. Energy level diagrams for the defect types in their corresponding empty and filled states are given below in **(h-k)**. For modest to large positive biases in (h), the dopant can be either neutral or positive depending on the competing $\Gamma_{T1-B}$ and $\Gamma_{T-T1}$ rates. If positive, the bands are locally bent downward (bending depicted as the blue curves *vs.* the black curves). For modest negative biases in (i), the dopant is neutral. In (j) and (k), the negative T2 always bends the bands upward (purple curves) relative to a defect free region (black curves). This enhances the current in filled states so that $\Gamma_{T-B}$ (purple: when the defect is present) is greater than $\Gamma_{T-B}$ (no defect). The opposite occurs in empty states. **(l)** Constant height and constant current **(m)** STM images of a T1 and T2 in the same 10×10 nm$^2$ area. **(n)** KPFM grid of the same area as (l,m) (Grid Dimensions = 25×25, $z_{rel}$ = 0.0 pm, $V_{range}$ = -1.8 - 1.0 V, and Osc. Amp = 50 pm). For the grid, the contact potential difference of the surface background has been subtracted. All scale bars are 5 nm.

In order to determine the volume concentration of the two near-surface species, their appearance in five large area frames (70×70 nm$^2$) was analyzed. Arsenic dopants were found to have an areal concentration of $(1.8\pm0.3)\times10^{11}$ atom·cm$^{-2}$ and T2s $(6\pm1)\times10^{10}$ atom·cm$^{-2}$. The maximum depth wherein dopants create assignable features in STM topography have been reported to range from 5 monolayers (ML)[35,39,43] to 36 ML.[40] The volume concentration based on these lower and upper bounds are $(2.6\pm0.4)\times10^{18}$ atom.cm$^{-3}$ (5 ML) and $(3.6\pm0.5)\times10^{17}$ atom.cm$^{-3}$ (36 ML) for arsenic dopants and $(1.0\pm0.2)\times10^{18}$ atom.cm$^{-3}$ (5 ML) and $(1.3\pm0.3)\times10^{17}$ atom.cm$^{-3}$ (36 ML) for T2s. Previous studies have reported that the As concentration near the surface is reduced during the sample preparation conditions applied here, from the bulk value of $1.5\times10^{19}$ atom.cm$^{-3}$ (See Methods-Sample Preparation). [44,45] The resulting dopant depleted layer extends ~70 nm into the bulk and has a As concentration of $1.0\times10^{18}$ atom.cm$^{-3}$ at the surface, [44,45] in agreement with the values observed here.

**Electrostatic Variation from Near-Surface Charges**

When probed, the charge state of the arsenic dopants can be neutral, positive, or negative depending on tip-surface distance and the applied bias which controls the competing filling and emptying rates from the tip and to the bulk, as reported in prior works.[35,46] Here, we explore conditions consistent with the positive and neutral charge state of the dopant.

Beginning with Figure 1d, taken at positive sample bias (V = +300 mV corresponding to empty sample states), the charge state of the dopant is, on average, neutral. Examining the corresponding energy level diagram in Figure 1h, when imaged in energy regimes above the onset of the bulk conduction band edge, the tip induced band bending (TIBB) at positive sample biases raises the dopant level above the bulk Fermi level ($E_{FS}$), but leaves it lower than the tip Fermi-level ($E_{FT}$). The charge state of the dopant depends on the competition among the filling rate from the tip ($\Gamma_{T-T1}$) and the emptying rate to the bulk ($\Gamma_{T1-B}$). If the tunneling rates to the dopant are tuned through adjustment of tip-sample separation or applied bias such that $\Gamma_{T1-B} < \Gamma_{T-T1}$, the dopant is rendered neutral with a single bound electron on average. However, for $\Gamma_{T1-B} > \Gamma_{T-T1}$, which would occur for modest biases between 0 V and the flat band condition, the dopant would be positive, in agreement with the presented electrostatic topography maps and spectroscopic shifts of our probe presented later. At 0 V, where there cannot be a net tunneling current, the dopant would be positively charged because of the upward band bending from the contact potential difference between the tip and sample [6] lifting the dopant level higher than both tip and sample Fermi-levels (Supporting Information, Figure S1c). The positive charge reduces the local band bending (blue lines) in the vicinity of the dopant.[25] Below the flat band condition, but above ~ -1.2 V, the dopant is again neutral. A constant-height STM image is presented in Figure 1e, with its corresponding energy level diagram in Figure 1i. The dopant becomes resonant with the bulk conduction band, and due to the narrower tunneling barrier between the dopant and the bulk conduction band, electrons tunnel to the dopant faster than the tip can extract them ($\Gamma_{B-T1} > \Gamma_{T1-T}$). The dopant opens up an additional channel for electrons to conduct from the tip to the conduction band, resulting in an increased conductivity over dopants compared to H-Si.[47] Below ~ -1.2 V, the dopant can become negative [40,41] or positive [47] depending on the dopant concentration at the surface and the relative magnitude of filling and emptying rates. A full energy level diagram progression over an arsenic atom from positive tip-sample biases to ~ -1.2 V is given in Supporting Information, Figure S1 for clarity.

A T2 charge defect is shown in Figure 1f where it appears as a region of reduced brightness in empty states, and in Figure 1g as a region of enhanced brightness in filled states. This appearance is consistent with the defect being negatively charged, as reported elsewhere.[35] In the qualitative empty states energy level diagram (Figure 1j), the upward band bending from the presence of the negatively charged T2 defect reduces the tunneling current from the tip to the conduction band (purple bands and purple $\Gamma_{T-B}$), resulting in the T2 showing up as a dark depression. [35] Without the T2 present, the bands are not altered (black bands and $\Gamma_{T-B}$). In filled states, the defect's bright contrast (Fig. 1g) indicates that it remains negatively charged, with the resulting local upward band bending (purple lines and purple Γ$_{B-T}$ in Fig. 1k) opening up a larger energy window for tunneling from the valence band to the tip.[35] These observations strongly suggest that the defect remains in the same negative charge state over the range of bias voltages examined here.

Both an arsenic and a T2 are shown together in the STM constant-height image Figure 1l, and the corresponding constant-current image Figure 1m. To map out the local electrostatic potential of the area, a 25×25 points grid was overlaid and constant-height *Δf(V)* curves were taken above each grid site. The *Δf(V)* spectra appear as parabolas with a maximum voltage V* representing the bias where the local contact potential difference has been nullified [15,18–20] (See Methods-KPFM Maps and Supporting Information, Figures S2,S3). An average V*, representing the electrostatic background, was calculated to be -0.49 V from averaging fifteen measurements over H-Si far from any charge-defects. Figure 1n presents the KPFM map of the extracted maxima (V*) mapping out the variation in the local electrostatic potential, with the background V* of H-Si subtracted. The upper left of the frame shows a dark area due to the presence of the arsenic dopant, indicating that at energies below the bulk Fermi level and above the flat band condition, it is positively charged reducing the contact potential difference in the area (bands are bent down as shown in Figure 1i). Conversely, the T2 gives rise to an enhancement in the contact potential difference in the lower right, in agreement with STM observations.[35]

**Dangling Bond Point-Probe**

A single DB was employed as a charge sensor. Figure 2a-h shows the progressive patterning [27–29] and erasure [30,31] of a single probe DB in the vicinity of an arsenic dopant. The lateral distance from the DB to the apparent center ranges from zero (directly on top of the arsenic dopant) to six surface lattice sites away as indicated in Fig. 2i (color-coded throughout frames a-h and the spectra in Figure 2l).

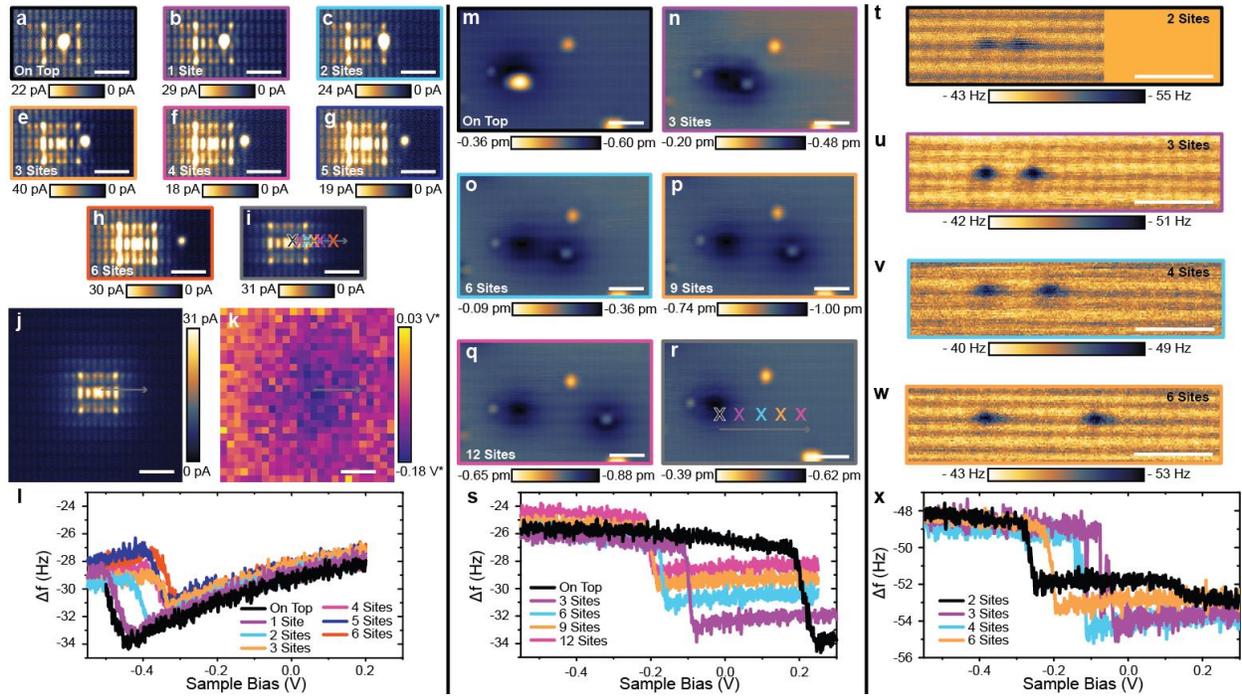

**Figure 2.** Probing charged species with a movable DB point-probe. **(a-h)** Constant height STM images of the DB being moved 0-6 lattice sites away from its initialization point, respectively, from a near-surface arsenic atom. ($z_{rel}$ = -250 pm and V = 300 mV) **(i)** Constant height STM image of the area before addition of a DB, with $\Delta f(V)$ locations marked with X's. ($z_{rel}$ = -200 pm and V = 250 mV) **(j)** Larger area constant height frame of (i). **(k)** KPFM grid of the same area as (j). (Grid Dimensions = 25×25 points, $z_{rel}$ = -200 pm, $V_{range}$ = -1.3 to 600 mV). The contact potential difference of the surface background has been subtracted out. **(l)** $\Delta f(V)$ spectroscopy taken on top of the DB for each lattice spacing, color coded with the positions in (i), as well as with the frames in (a-h) ($z_{rel}$ = -350 pm). **(m-q)** Constant current STM images (V = 1.3 V and I = 50 pA) of the DB being moved 0,3,6,9, and 12 lattice sites away from its initialization point, respectively, from a T2. **(r)** Constant current STM image marking the $\Delta f(V)$ locations. **(s)** $\Delta f(V)$ spectroscopy taken on top of the DB at the listed lattice spacing's ($z_{rel}$ = -300 pm). **(t-w)** Constant

height AFM images of a DB 2, 3, 4, and 6 lattice sites away from another DB ($z_{rel}$ = -300 pm, V = 0 V). **(x)** *Δf(V)* spectroscopy taken on top of the left DB at the listed lattice spacing's ($z_{rel}$ = -300 pm). All scale bars are 2 nm.

A KPFM map taken of the area before the addition of the DB probe is shown in Fig. 2k, with its corresponding STM image in Fig. 2j. The map shows a local dark depression at the location of the dopant. The *Δf(V)* spectra for the different lattice positions are displayed in Figure 2l with the black curve being closest to the arsenic dopant and the dark orange farthest away. As mentioned earlier, the DB is capable of having three different charge states with two charge transition levels. A change of the DB's charge state is manifested as a single-electron-charge transition step [6,32–34] in the AFM *Δf(V)* spectroscopy, with the DB switching from negative (doubly occupied) for bias voltages to the right of the step in the spectra, to neutral for bias voltages on the left of the step (see Supporting Information, Figure S3). Over the distance of 2.3 nm that the DB was moved, the (0/-) charge transition step is seen to shift from -0.44±0.01 V to -0.27±0.01 V, with the bias at which the step occurs being more negative closer to the arsenic. This picture is once again consistent with a positive charge at the dopant atom bending the bands locally, requiring the application of a larger negative tip-sample bias to withdraw an electron from the DB.

This same kind of DB point-probe analysis was also performed for a T2, shown in the constant-current STM images of Figure 2m-q, with the spacing ranging from zero to twelve lattice sites over a 4.6 nm distance. Figure 2r marks the locations before the DB was added. Examining the *Δf(V)* curves in Figure 2s, a reversed trend for the shift is observed when compared to the positively charged arsenic. Now the charge transition of the DB is shifted to

+0.21 V for the nearest spectra (black curve) and -0.20 V for farthest (pink curve). This is indicative of a negatively charged species.

A final analysis was done by moving a probe DB away from a second DB as shown in the AFM constant height images of Figure 2t-w, where it was moved two, three, four, and six lattice sites away (1.9 nm distance total), respectively. In ref. [6] we showed a related analysis using the charge transition states of DBs assembled in fixed arrangements. For large separations between the DBs the distance-dependent shift of the probe DB's charge transition level clearly indicates the negative charge located at the static DB. Unlike the prior two charge species, the $\Delta f(V)$ curves in Figure 2x completely change behavior for small separations. The black curve in Figure 2x for the closest-spaced pair (two lattice sites) exhibits two charge transition steps. The reason for this behavior is the shared occupation for the created DB-DB pair.[6] The left step at -0.24 V corresponds to the case that only one electron is present in the pair (located at one of the DBs). Sweeping to less negative bias voltages the second DB becomes charged, creating the second charge transition step at -0.03 V. With increasing distance within the pair the two DBs act independently, and individual single charge transitions are observed that are mutually at less negative values (purple curve in Fig 2x). As the DBs are positioned progressively farther apart with four (blue curve) and six (orange curve) lattice sites between them respectively, their mutual shifting effect on each other lessens and the charge transition approaches that of an isolated DB again.

**Fitting Shifts of the Dangling Bond Probe**

The values of the electrostatic shifts of the (0/-) charge transition voltage of the probe DB were extracted for the arsenic, T2, and the DB interactions (See Supporting Information, Figure S4 for corresponding IV curves for Fig. 2l,s,x with vertical color-coded lines marking the extracted charge transition shifts). Each shift was corrected for tip induced band bending (TIBB)[48–50] (See Methods-Tip Induced Band Bending), plotted as a function of distance in Figure 3, and fit with a linearized form of the screened Coulomb energy equation:

$$\ln(r\,U(r)) = \ln\left(\frac{e^2}{4\pi\varepsilon_0\varepsilon}\right) - \frac{r}{L_{TF}}$$

where $e$ is the elementary charge, $\varepsilon$ the effective dielectric constant, $r$ the distance between the DB and the localized charge, and $L_{TF}$ the Thomas-Fermi screening length (See Methods-Fitting for details). The DB-DB shifts (orange) are in the same height plane, but the near-surface charges also have a depth component that can be extracted as well ($r = \sqrt{(\text{lateral distance})^2 + (\text{depth})^2}$).

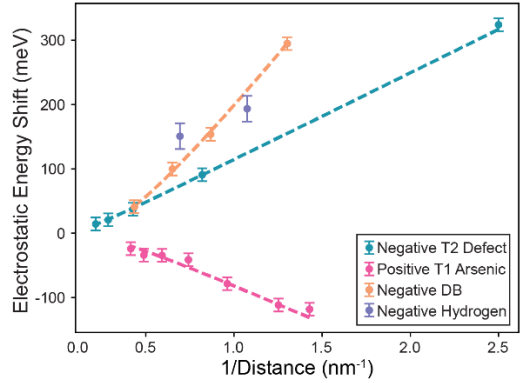

| Charge Defect | Dielectric | Depth (nm) | $L_{TF}$ (nm) |
|---|---|---|---|
| T1 Arsenic | 9.7 ± 2.5 | 0.8 | 2.1 ± 0.7 |
| T2 | 10.6 ± 0.5 | 0.4 | 5.9 ± 0.6 |
| DB | 4.1 ± 0.2 | 0.0 | 1.8 ± 0.1 |

**Figure 3.** Fitting DB charge transition shifts. Shifts of the charge transition step for the probe DB as a function of distance from the T1+ (pink curve), T2- (blue curve), a second DB- (orange curve), and a physisorbed hydrogen (purple). Tip induced band bending was subtracted from the experimental data (see Methods-Tip Induced Band Bending). Error bars correspond to the read-out error of the electrostatic energy shift, taken to be ±10 mV for T1, T2, and DB cases, and ±20 mV for hydrogen. Dashed lines are an orthogonal distance regression fit of the data to the linearized form of the screened Coulomb equation. Errors of the dielectric constant and screening length correspond to the standard error (See Methods-Fitting).

Other work has been able to estimate these parameters from fitting the spatially resolved spectral shift of the conduction band edge to a dielectric screened Coulomb potential,[41,49] as well as through comparison to simulated STM images created using a tight-binding framework.[40] When compared to the tight-binding framework,[40] the method employed here has the advantage of being more general in that *a priori* knowledge of the identity of the defect is not needed to extract the information, nor is atomistic modeling. Our method has parallels to STM fitting of the shift in the conduction band edge [41,49] but can access bias regimes inaccessible to STM.

The extracted values using our method are summarized in the table in Figure 3. In our fitting procedure, depth has no associated error as it was incremented in the fit code with $\varepsilon$ and $L_{TF}$ being fit as free parameters (See Methods-Fitting). Two points extracted from a fourth experiment involving a physisorbed hydrogen atom, discussed in Supporting Information, Negative Hydrogen, are displayed in purple, but were not fitted due to an insufficient number of data points.

Both near-surface defect types reveal dielectric constants close to the established value for bulk silicon (11.7), whereas the dielectric constant determined from the DB-DB case is closer to the expected average for the silicon/vacuum interface of 6.35.[41] The screening length varies by several nanometers among the experiments presented here. These variations do not seem to be specific to the types of species, with the screening length extracted in a similar DB experiment in ref. [6] being 5 nm, (1.8±0.1 nm here). These variations further emphasize that the carrier density is not uniform across the sample surface, highlighting the need for local characterization of the electrostatic environment.

**Screening for Homogeneous Areas**

For atomic devices fabricated on the surface, having a homogeneous electrostatic environment with predictable DB charge transition energies is vital to ensure consistent device operation. It has been established in this work that KPFM maps are capable of resolving small changes in the local electrostatic environment of H-Si. In Figure 4, this technique is used to pre-screen an area for electrostatic homogeneity.

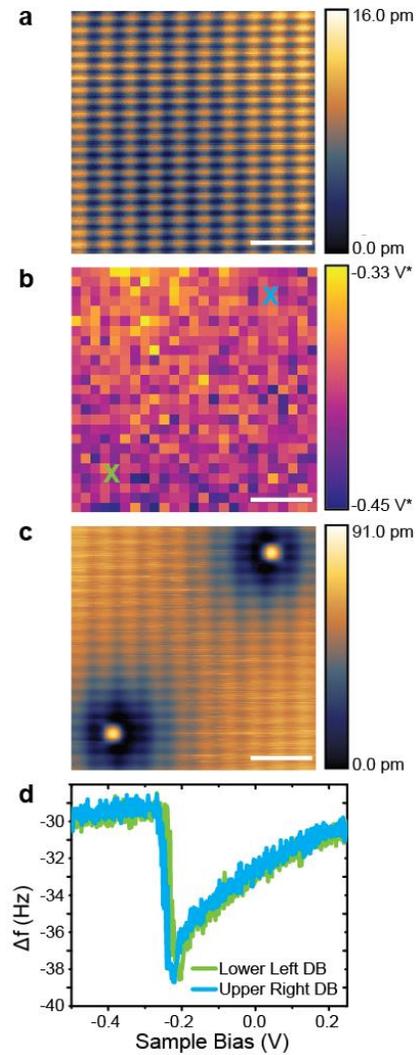

**Figure 4.** Screening for a uniform electrostatic background. **(a)** Constant current empty states (V = 1.3 V and I = 50 pA) STM image of a clean H-Si area. **(b)** KPFM grid taken of the same area. Two areas with similar contrast are marked with color coded X's for patterning of a DB (Grid Dimensions = 25×25 points, $z_{rel}$ = 0.0 pm, $V_{range}$ = -1.3 - 1.0 V). **(c)** Constant current image once the two DBs were patterned at opposite corners. **(d)** *Δf(V)* spectroscopy taken on top of the two DBs showing similar charge transition voltages. ($z_{rel}$ = -350 pm). All scale bars are 2 nm.

Figure 4a shows a STM image of a 8×8 nm area of H-Si. A KPFM map taken of this area is displayed in Figure 4b, and two areas with a similar contact potential difference value were selected (marked by the colored crosses). Figure 4c shows the area after the creation of the two DBs at the marked locations. *Δf(V)* curves taken over the DBs are shown in 4d, showing similar charge transition energies for both DBs.

**Identity of the T2 Charge Feature**

As mentioned earlier, the identity of T2s has been assigned inconclusively throughout the recent literature. References [39] and [42] hypothesized them to be boron contamination due to their acceptor like behavior. Reference [35] noted that this was unlikely because boron contamination at the concentrations reported is highly unlikely in commercial-grade wafers and instead proposed them to be negatively charged arsenic dopants.[35] The corresponding two-electron state has been reported for both phosphorous [9,51,52] and arsenic [41,53,54] donors, but that species is weakly bound with a binding energy only a few meV below the silicon conduction band edge for bulk donors. While this state exists at filled state biases and is observed as a peak in dI/dV spectroscopy,[41] it cannot exist at large empty state biases [54] or elevated sample temperatures as any electron potentially bound in that state would readily delocalize to the conduction band. Observations of T2 at room temperature as well as at empty state biases [43] are therefore inconsistent with the negative donor assignment.

We propose that the T2- features are more likely to originate from either negatively charged interstitial hydrogen atoms or a negative hydrogen-vacancy complex. We base our assignment on the sample fabrication process in a hydrogen-rich atmosphere and that the observed T2s are negatively charged. During preparation of the H-Si(100)-2×1 surface (See Methods-Sample

Preparation), the silicon sample is rapidly heated to 1250 °C and then annealed at a temperature of 330 °C for several minutes in an atmosphere of atomic hydrogen.[44,55] The elevated temperatures during our sample termination procedure is known to allow hydrogen atoms to readily enter the crystal [56–58] and secondary-ion mass spectrometry revealed that hydrogen is able to penetrate several microns into the silicon wafer, depending on parameters such as doping type, concentration, and growth method.[56,58] In the very near surface region, the concentration of hydrogen atoms can be comparable [56,58] to the doping level of the silicon crystal used in our present work.[44,45] The hydrogen atoms can occupy various sites depending on their charge state.[57] They may be at defects,[56–58] form complexes with silicon [59–63] or with dopant atoms, or they may reside interstitially.[57]

The charge state of the suspected hydrogen atom or hydrogen-silicon complexes is determined by the Fermi-level of the bulk crystal, which in the present case is in a regime where such species would be negative.[57,59–61] For hydrogen, the negative ions are reported to sit at the tetrahedral interstitial sites of the silicon lattice and exhibit a negative-U character,[57] that is, the second bound electron is held more strongly than the first due to a repositioning of the hydrogen within the lattice after gaining a negative charge.[57,64] Even lone hydrogen atoms that remain on the surface after the creation of a DB [30] can possess a negative charge. Supporting Information, Negative Hydrogen shows how DBs are affected by the negative charge of a nearby H atom. While a direct comparison between the two forms, on-surface *vs.* bulk-trapped hydrogen, is impossible to be drawn due to the different local environments, the resemblance of their effect on DBs is striking. To determine conclusively whether the bulk-bound negative species are lone hydrogen atoms or a hydrogen-vacancy complex will require extensive high-level atomistic modeling of the various candidates. However, the stability of the negative charge state of hydrogen atoms and the abundance of hydrogen in the near surface region are strong indications for our proposal that the T2- features involve hydrogen atoms.

**Conclusion and Outlook**

In this work, we used KPFM mapping to show surface electrostatic variation on a nm length scale caused by two oppositely charged near-surface species. We then went on to explore three charged species on the H-Si(100)-2×1 surface. In agreement with the literature, we assigned the T1 species to positively charged arsenic donors. The T2 species were shown to be conclusively negative in character, and we proposed that they may be negatively charged hydrogen atoms or negative hydrogen-vacancy complexes formed during sample preparation.

The quantum dot probe, reminiscent of scanning quantum dot microscopy, was realized through the sequential patterning and capping of a "moveable" DB. We showed that a charged species was able to shift the charge transition voltage of the sensitive quantum dot probe DB with the expected reciprocal relation to absolute distance. Through analysis and fitting of the observed shifts of the charge transition levels in the on-surface DB point-probe, we were able to determine the dielectric constant, the screening length, and experimentally determine the depth of the species. The DB-DB analysis also allowed extraction of an experimentally determined distance where the DB pair occupation changes, confirming less direct measures and with important implications for DB-based device applications.[5–8]

Our approach of placing the single atom quantum dot directly on the surface of interest provides the highest spatial resolution of lateral potential variations as measured by scanning quantum dot microscopy to date. Direct comparison of our observations with complementary KPFM maps highlights the impact that charged species have on the electrostatic landscape of a surface. The various species studied and their different electrostatic impact over several nanometers underlines the necessity to carefully select areas for atom-scale device applications.

**Methods**

**Measurement System Setup:** Experiments were performed in a commercial (Scienta-Omicron) qPlus AFM [65,66] system operating at 4.5 K. Nanonis scanning probe control electronics and Specs software were used to acquire both the STM and AFM data. For all constant-height images, $z_{rel} = 0$ pm corresponds to the relative tip elevation defined by the STM set points of I = 50 pA and V = -1.8 V on top of a hydrogen-terminated silicon atom. The tuning fork had a resonance frequency of 28.2 kHz, with a quality factor of ~17, 000. To minimize drift during the KPFM maps, the tip was left to settle for a minimum of 12 hours before starting a map.

**Sample Preparation:** Highly arsenic-doped (~1.5×10$^{19}$ atom.cm$^{-3}$) Si(100) was used. Overnight degassing in ultra-high vacuum (UHV) was done at ~600 °C for 12 hours to prepare the sample for flashing. Once degassed, a series of resistive flash anneals to 1250 °C to remove oxide were done.[44,45,55] The series of resistive flash anneals has been shown to reduce surface dopant density, creating a region depleted of dopants ~70 nm below the sample surface with a donor concentration ~10$^{18}$ atom.cm$^{-3}$.[44,45] The final step to terminate with hydrogen was done by holding the Si substrate at 330 °C for two minutes while molecular hydrogen (pressure = 10$^{-6}$ Torr) was cracked with a 1600 °C tungsten filament.[44,55]

**Tip Preparation:** qPlus-style quartz tuning fork AFM sensors were used for all experiments.[65,66] Once in UHV, tips were prepared by field evaporating the apex clean in a field ion microscope (FIM).[67] The tip was then sharpened by a FIM nitrogen etching process.[67] Final *in-situ* conditioning consisted of controlled contacts on patches of desorbed silicon.[68]

Creating and Erasing DBs: To create a DB, a sharp tip is positioned over the target atom at 1.3 V and 50 pA, and pulses of 2.0 - 2.5 V for 10 ms are applied until the hydrogen is removed.[27–29,69] Sometimes the removed hydrogen atom functionalizes the tip apex. A hydrogen atom terminated tip is immediately made obvious by enhanced topographic corrugations.[27,30] Such a tip can then be used to erase a DB.[30,31] This is done by bringing the functionalized tip at 0 V mechanically towards the DB to induce a H-Si covalent bond and the passivation of the DB.[27,30,31]

KPFM Maps: The tip is left to settle overnight to minimize piezo creep during the experiment. Nanonis software supports a grid experiment feature, which is used to automate the acquisition of a $\Delta f(V)$ curve for every point in a defined grid. The presented map in Figure 1n took ~2 hours to acquire. Python code was used to fit a parabolic curve of the form $y=Ax^2+Bx+C$ to each $\Delta f(V)$ and extract the maximum (V*) of the fit parabola.[15,33] An example of the fit for a single point in Figure 2n is shown in Supporting Information, Figure S2. These V* maxima are extracted for every curve, and then plotted as a KPFM map. Taking KPFM maps directly over DBs using the same voltage range as used for probing near-surface charges was avoided to prevent damage to the tip as a result of high tunneling current. KPFM maps can be taken over DBs, but using more modest ranges or conservative heights. An example of a KPFM curve over a DB with parabolas fit to the negative and neutral charge states is shown in Supporting Information, Figure S3.

**Tip Induced Band Bending:** During scanning probe imaging, a contact potential difference exists between the tip and sample due to their different work functions. For the tungsten tip the work function is estimated to be 4.5 eV, while for the n-doped silicon sample with a dopant depleted layer concentration of $10^{18}$ atom.cm$^{-3}$ (at low temperature) is estimated at 4.1 eV. This difference creates band-bending locally under the tip apex that can shift the charge transition levels of the dangling bonds (tip-induced band bending, TIBB). While the contact potential difference is approximately constant, the TIBB predictably changes with varied tip-sample separation, and applied tip-sample bias. This has implications for the energy levels extracted in Supporting Information, Figure S4 from the *Δf(V)* spectra. The energy level shifts extracted should have the TIBB factored in to give a band-bending free picture of the true shifts. The TIBB was calculated using a 3D finite-element Poisson equation solver.[6,48] For the calculation, a work function difference between tip and sample of 0.4 eV was assumed, a tip radius of 10 nm, a tip-sample height of $z_{rel}$ = -300 or -350 pm (see figure captions), and a donor concentration of $10^{18}$ atom.cm$^{-3}$ at the surface, monotonically increasing to $2\times10^{19}$ atom.cm$^{-3}$ in the bulk over a range of approximately 100 nm.

**Fitting:** The screened Coulomb equation (Equation 1) was fit to the experimental data in a linearized form (Equation 2) with an orthogonal distance regression method:

$$U(r) = \frac{e^2}{4\pi\varepsilon_0 \varepsilon r} e^{-r/L_{TF}} \quad (1)$$

$$\ln(r\, U(r)) = \ln\left(\frac{e^2}{4\pi\varepsilon_0 \varepsilon}\right) - \frac{r}{L_{TF}} \quad (2)$$

In the DB-DB experiment, the two free parameters of $\varepsilon$ and $L_{TF}$ were fit independently and the estimated experimental read-out error (±10 mV) was factored in by using an orthogonal distance regression python algorithm. For the subsurface arsenic and T2, depth was incremented in steps of 0.1 nm from zero to an assumed maximum detectable depth of 5 nm.[40] For every depth increment, ε, $L_{TF}$, and the associated standard error for the two parameters was extracted. The values of depth, ε, and $L_{TF}$ presented correspond to the parameters that minimized the combined standard errors.

**Corresponding Author**

Correspondence and requests for materials should be addressed to taleana@ualberta.ca; rwolkow@ualberta.ca.

**Funding Sources**

We thank NRC-NRC, NSERC, QSi, Compute Canada, and Alberta Innovates for their financial support.

**Acknowledgements**

We thank M. Cloutier, and M. Salomons for their technical expertise. We thank W. Vine for helpful comments and critiques of the manuscript.

## References

(1) Khajetoorians, A. A.; Wiebe, J.; Chilian, B.; Wiesendanger, R. Realizing All-Spin–Based Logic Operations Atom by Atom. *Science* **2011**, *332*, 1062–1064.

(2) Heinrich, A. J.; Lutz, C. P.; Gupta, J. A.; Eigler, D. M. Molecule Cascades. *Science* **2002**, *298*, 1381–1387.

(3) Joachim, C.; Gimzewski, J. K.; Aviram, A. Electronics Using Hybrid-Molecular and Mono-Molecular Devices. *Nature* **2000**, *408*, 541.

(4) Ratner, M. A Brief History of Molecular Electronics. *Nat. Nanotechnol.* **2013**, *8*, 378.

(5) Yengui, M.; Duverger, E.; Sonnet, P.; Riedel, D. A Two-Dimensional ON/OFF Switching Device Based on Anisotropic Interactions of Atomic Quantum Dots on Si(100):H. *Nat. Commun.* **2017**, *8*, 2211.

(6) Huff, T.; Labidi, H.; Rashidi, M.; Livadaru, L.; Dienel, T.; Achal, R.; Vine, W.; Pitters, J.; Wolkow, R. A. Binary Atomic Silicon Logic. *Nat. Electron.* **2018**, *1*, 636–643.

(7) Wyrick, J.; Wang, X.; Namboodiri, P.; Schmucker, S. W.; Kashid, R.; Silver, R. Atom-by-Atom Construction of a Cyclic Artificial Molecule in Silicon. *Nano Lett.* **2018**.

(8) Schofield, S. R.; Studer, P.; Hirjibehedin, C. F.; Curson, N. J.; Aeppli, G.; Bowler, D. R. Quantum Engineering at the Silicon Surface Using Dangling Bonds. *Nat. Commun.* **2013**, *4*, 1649.

(9) Fuechsle, M.; Miwa, J. A.; Mahapatra, S.; Ryu, H.; Lee, S.; Warschkow, O.; Hollenberg, L. C. L.; Klimeck, G.; Simmons, M. Y. A Single-Atom Transistor. *Nat. Nanotechnol.* **2012**, *7*, 242–246.

(10) Fresch, B.; Bocquel, J.; Rogge, S.; Levine, R. D.; Remacle, F. A Probabilistic Finite State Logic Machine Realized Experimentally on a Single Dopant Atom. *Nano Lett.* **2017**, *17*, 1846–1852.

(11) Pica, G.; Lovett, B. W. Quantum Gates with Donors in Germanium. *Phys. Rev. B* **2016**, *94*, 205309.

(12) Hollenberg, L. C. L.; Dzurak, A. S.; Wellard, C.; Hamilton, A. R.; Reilly, D. J.; Milburn, G. J.; Clark, R. G. Charge-Based Quantum Computing Using Single Donors in Semiconductors. *Phys. Rev. B* **2004**, *69*, 113301.

(13) Sigillito, A. J.; Tyryshkin, A. M.; Schenkel, T.; Houck, A. A.; Lyon, S. A. All-Electric Control of Donor Nuclear Spin Qubits in Silicon. *Nat. Nanotechnol.* **2017**, *12*, 958–962.

(14) Gohlke, D.; Mishra, R.; Restrepo, O. D.; Lee, D.; Windl, W.; Gupta, J. Atomic-Scale Engineering of the Electrostatic Landscape of Semiconductor Surfaces. *Nano Lett.* **2013**, *13*, 2418–2422.

(15) Mohn, F.; Gross, L.; Moll, N.; Meyer, G. Imaging the Charge Distribution within a Single Molecule. *Nat. Nanotechnol.* **2012**, *7*, 227.

(16) Albrecht, F.; Repp, J.; Fleischmann, M.; Scheer, M.; Ondráek, M.; Jelínek, P. Probing Charges on the Atomic Scale by Means of Atomic Force Microscopy. *Phys. Rev. Lett.* **2015**, *115*, 76101.

(17) Weymouth, A. J.; Wutscher, T.; Welker, J.; Hofmann, T.; Giessibl, F. J. Phantom Force Induced by Tunneling Current: A Characterization on Si(111). *Phys. Rev. Lett.* **2011**, *106*, 226801.

(18) Schuler, B.; Liu, S.-X.; Geng, Y.; Decurtins, S.; Meyer, G.; Gross, L. Contrast Formation in Kelvin Probe Force Microscopy of Single π-Conjugated Molecules. *Nano Lett.* **2014**, *14*, 3342–3346.

(19) Barth, C.; Henry, C. R. Surface Double Layer on (001) Surfaces of Alkali Halide Crystals: A Scanning Force Microscopy Study. *Phys. Rev. Lett.* **2007**, *98*, 136804.

(20) Barth, C.; Foster, A. S.; Henry, C. R.; Shluger, A. L. Recent Trends in Surface Characterization and Chemistry with High-Resolution Scanning Force Methods. *Adv. Mater.* **2010**, *23*, 477–501.

(21) Wagner, C.; Green, M. F. B.; Leinen, P.; Deilmann, T.; Krüger, P.; Rohlfing, M.; Temirov, R.; Tautz, F. S. Scanning Quantum Dot Microscopy. *Phys. Rev. Lett.* **2015**, *115*, 026101.

(22) Ondráček, M.; Hapala, P.; Švec, M.; Jelínek, P. Imaging Charge Distribution Within Molecules by Scanning Probe Microscopy-Kelvin Probe Force Microscopy: From Single Charge Detection to Device Characterization; Sadewasser, S., Glatzel, T., Eds.; Springer International Publishing: Cham, 2018; pp 499–518.

(23) Haider, M. B.; Pitters, J. L.; DiLabio, G. A.; Livadaru, L.; Mutus, J. Y.; Wolkow, R. A. Controlled Coupling and Occupation of Silicon Atomic Quantum Dots at Room Temperature. *Phys. Rev. Lett.* **2009**, *102*, 046805.

(24) Rashidi, M.; Lloyd, E.; Huff, T. R.; Achal, R.; Taucer, M.; Croshaw, J. J.; Wolkow, R. A. Resolving and Tuning Carrier Capture Rates at a Single Silicon Atom Gap State. *ACS Nano* **2017**, *11*, 11732–11738.

(25) Taucer, M.; Livadaru, L.; Piva, P. G.; Achal, R.; Labidi, H.; Pitters, J. L.; Wolkow, R. A. Single-Electron Dynamics of an Atomic Silicon Quantum Dot on the H-Si(100) - (2x1) Surface. *Phys. Rev. Lett.* **2014**, *112*, 256801.

(26) Bellec, A.; Chaput, L.; Dujardin, G.; Riedel, D.; Stauffer, L.; Sonnet, P. Reversible Charge Storage in a Single Silicon Atom. *Phys. Rev. B* **2013**, *88*, 241406.

(27) Achal, R.; Rashidi, M.; Croshaw, J.; Churchill, D.; Taucer, M.; Huff, T.; Cloutier, M.; Pitters, J.; Wolkow, R. A. Lithography for Robust and Editable Atomic-Scale Silicon Devices and Memories. *Nat. Commun.* **2018**, *9*, 2778.

(28) Shen, T.-C.; Wang, C.; Abeln, G. C.; Tucker, J. R.; Lyding, J. W.; Avouris, P.; Walkup, R. E. Atomic-Scale Desorption Through Electronic and Vibrational Excitation Mechanisms. *Science* **1995**, *268*, 1590–1592.

(29) Lyding, J. W.; Shen, T. -C.; Hubacek, J. S.; Tucker, J. R.; Abeln, G. C. Nanoscale Patterning and Oxidation of H-passivated Si(100)-2×1 Surfaces with an Ultrahigh Vacuum Scanning Tunneling Microscope. *Appl. Phys. Lett.* **1994**, *64*, 2010–2012.

(30) Huff, T. R.; Labidi, H.; Rashidi, M.; Koleini, M.; Achal, R.; Salomons, M. H.; Wolkow, R. A. Atomic White-Out: Enabling Atomic Circuitry through Mechanically Induced Bonding of Single Hydrogen Atoms to a Silicon Surface. *ACS Nano* **2017**, *11*, 8636–8642.

(31) Pavliček, N.; Majzik, Z.; Meyer, G.; Gross, L. Tip-Induced Passivation of Dangling Bonds on Hydrogenated Si(100)-2×1. *Appl. Phys. Lett.* **2017**, *111*, 053104.

(32) Steurer, W.; Fatayer, S.; Gross, L.; Meyer, G. Probe-Based Measurement of Lateral Single-Electron Transfer Between Individual Molecules. *Nat. Commun.* **2015**, *6*, 8353.

(33) Gross, L.; Mohn, F.; Liljeroth, P.; Repp, J.; Giessibl, F. J.; Meyer, G. Measuring the Charge State of an Adatom with Noncontact Atomic Force Microscopy. *Science* **2009**, *324*, 1428–1431.

(34) Steurer, W.; Repp, J.; Gross, L.; Scivetti, I.; Persson, M.; Meyer, G. Manipulation of the Charge State of Single Au Atoms on Insulating Multilayer Films. *Phys. Rev. Lett.* **2015**, *114*, 036801.

(35) Sinthiptharakoon, K.; Schofield, S. R.; Studer, P.; Brázdová, V.; Hirjibehedin, C. F.; Bowler, D. R.; Curson, N. J. Investigating Individual Arsenic Dopant Atoms in Silicon Using Low-Temperature Scanning Tunnelling Microscopy. *J. Phys. Condens. Matter* **2014**, *26*, 12001.

(36) Spadafora, E. J.; Berger, J.; Mutombo, P.; Telychko, M.; Švec, M.; Majzik, Z.; McLean, A. B.; Jelínek, P. Identification of Surface Defects and Subsurface Dopants in a Delta-Doped System Using Simultaneous Nc-AFM/STM and DFT. *J. Phys. Chem. C* **2014**, *118*, 15744–15753.

(37) Yap, T. L.; Kawai, H.; Neucheva, O. A.; Wee, A. T. S.; Troadec, C.; Saeys, M.; Joachim, C. Si(100)-2×1-H Dimer Rows Contrast Inversion in Low-Temperature Scanning Tunneling Microscope Images. *Surf. Sci.* **2015**, *632*, L13–L17.

(38) Boland, J. J. Structure of the H-Saturated Si(100) Surface. *Phys. Rev. Lett.* **1990**, *65*, 3325–3328.

(39) Liu, L.; Yu, J.; Lyding, J. W. Subsurface Dopant-Induced Features on the Si (100) 2x1: H Surface: Fundamental Study and Applications. *IEEE Trans. Nanotechnol.* **2002**, *99*, 176–183.

(40) Usman, M.; Bocquel, J.; Salfi, J.; Voisin, B.; Tankasala, A.; Rahman, R.; Simmons, M. Y.; Rogge, S.; Hollenberg, L. C. L. Spatial Metrology of Dopants in Silicon with Exact Lattice Site Precision. *Nat. Nanotechnol.* **2016**, *11*, 763.

(41) Salfi, J.; Mol, J. A.; Rahman, R.; Klimeck, G.; Simmons, M. Y.; Hollenberg, L. C. L.; Rogge, S. Spatially Resolving Valley Quantum Interference of a Donor in Silicon. *Nat. Mater.* **2014**, *13*, 605–610.

(42) Liu, L.; Yu, J.; Lyding, J. W. Atom-Resolved Three-Dimensional Mapping of Boron Dopants in Si(100) by Scanning Tunneling Microscopy. *Appl. Phys. Lett.* **2001**, *78*, 386–388.

(43) Piva, P. G.; DiLabio, G. A.; Livadaru, L.; Wolkow, R. A. Atom-Scale Surface Reactivity Mediated by Long-Ranged Equilibrium Charge Transfer. *Phys. Rev. B* **2014**, *90*, 155422.

(44) Pitters, J. L.; Piva, P. G.; Wolkow, R. A. Dopant Depletion in the Near Surface Region of Thermally Prepared Silicon (100) in UHV. *J. Vac. Sci. Technol. B* **2012**, *30*, 21806.

(45) Labidi, H.; Taucer, M.; Rashidi, M.; Koleini, M.; Livadaru, L.; Pitters, J.; Cloutier, M.; Salomons, M.; Wolkow, R. A. Scanning Tunneling Spectroscopy Reveals a Silicon Dangling Bond Charge State Transition. *New J. Phys.* **2015**, *17*, 073023.

(46) Voisin, B.; Salfi, J.; Bocquel, J.; Rahman, R.; Rogge, S. Spatially Resolved Resonant Tunneling on Single Atoms in Silicon. *J. Phys. Condens. Matter* **2015**, *27*, 154203.

(47) Rashidi, M.; Burgess, J. A. J.; Taucer, M.; Achal, R.; Pitters, J. L.; Loth, S.; Wolkow, R. A. Time-Resolved Single Dopant Charge Dynamics in Silicon. *Nat. Commun.* **2016**, *7*, 13258.

(48) Ryan, P. M.; Livadaru, L.; DiLabio, G. A.; Wolkow, R. A. Organic Nanostructures on Hydrogen-Terminated Silicon Report on Electric Field Modulation of Dangling Bond Charge State. *J. Am. Chem. Soc.* **2012**, *134*, 12054–12063.

(49) Teichmann, K.; Wenderoth, M.; Loth, S.; Ulbrich, R. G.; Garleff, J. K.; Wijnheijmer, A. P.; Koenraad, P. M. Controlled Charge Switching on a Single Donor with a Scanning Tunneling Microscope. *Phys. Rev. Lett.* **2008**, *101*, 76103.

(50) Feenstra, R. M.; Dong, Y.; Semtsiv, M. P.; Masselink, W. T. Influence of Tip-Induced Band Bending on Tunnelling Spectra of Semiconductor Surfaces. *Nanotechnology* **2007**, *18*, 44015.

(51) Ramdas, A. K.; Rodriguez, S. Spectroscopy of the Solid-State Analogues of the Hydrogen Atom: Donors and Acceptors in Semiconductors. *Reports Prog. Phys.* **1981**, *44*, 1297.

(52) Tankasala, A.; Salfi, J.; Bocquel, J.; Voisin, B.; Usman, M.; Klimeck, G.; Simmons, M. Y.; Hollenberg, L. C. L.; Rogge, S.; Rahman, R. Two-Electron States of a Group-V Donor in Silicon from Atomistic Full Configuration Interactions. *Phys. Rev. B* **2018**, *97*, 195301.

(53) Rahman, R.; Lansbergen, G. P.; Verduijn, J.; Tettamanzi, G. C.; Park, S. H.; Collaert, N.; Biesemans, S.; Klimeck, G.; Hollenberg, L. C. L.; Rogge, S. Electric Field Reduced Charging Energies and Two-Electron Bound Excited States of Single Donors in Silicon. *Phys. Rev. B* **2011**, *84*, 115428.

(54) Salfi, J.; Voisin, B.; Tankasala, A.; Bocquel, J.; Usman, M.; Simmons, M. Y.; Hollenberg, L. C. L.; Rahman, R.; Rogge, S. Valley Filtering in Spatial Maps of Coupling between Silicon Donors and Quantum Dots. *Phys. Rev. X* **2018**, *8*, 31049.

(55) Boland, J. J. Scanning Tunnelling Microscopy of the Interaction of Hydrogen with Silicon Surfaces. *Adv. Phys.* **1993**, *42*, 129–171.

(56) Sopori, B. L.; Deng, X.; Benner, J. P.; Rohatgi, A.; Sana, P.; Estreicher, S. K.; Park, Y. K.; Roberson, M. A. Hydrogen in Silicon: A Discussion of Diffusion and Passivation Mechanisms. *Sol. Energy Mater. Sol. Cells* **1996**, *41–42*, 159–169.

(57) Herring, C.; Johnson, N. M.; de Walle, C. G. Energy Levels of Isolated Interstitial Hydrogen in Silicon. *Phys. Rev. B* **2001**, *64*, 125209.

(58) Johnson, N. M.; Herring, C.; Chadi, D. J. Interstitial Hydrogen and Neutralization of Shallow-Donor Impurities in Single-Crystal Silicon. *Phys. Rev. Lett.* **1986**, *56*, 769–772.

(59) Roberson, M. A.; Estreicher, S. K. Vacancy and Vacancy-Hydrogen Complexes in Silicon. *Phys. Rev. B* **1994**, *49*, 17040–17049.

(60) Park, Y. K.; Estreicher, S. K.; Myles, C. W.; Fedders, P. A. Molecular-Dynamics Study of the Vacancy and Vacancy-Hydrogen Interactions in Silicon. *Phys. Rev. B* **1995**, *52*, 1718–1723.

(61) Deák, P.; Heinrich, M.; Snyder, L. C.; Corbett, J. W. Hydrogen-Related Vibrations in Crystalline Silicon. *Mater. Sci. Eng. B* **1989**, *4*, 57–62.

(62) Deák, P.; Snyder, L. C.; Heinrich, M.; Ortiz, C. R.; Corbett, J. W. Hydrogen Complexes and Their Vibrations in Undoped Crystalline Silicon. *Phys. B Condens. Matter* **1991**, *170*, 253–258.

(63) Corbett, J. W.; Lindström, J. L.; Pearton, S. J. Hydrogen in Silicon. *MRS Proc.* **1987**, *104*, 229.

(64) Watkins, G. D.; Troxell, J. R. Negative-U Properties for Point Defects in Silicon. *Phys. Rev. Lett.* **1980**, *44*, 593–596.

(65) Giessibl, F. J. High-Speed Force Sensor for Force Microscopy and Profilometry Utilizing a Quartz Tuning Fork. *Appl. Phys. Lett.* **1998**, *73*, 3956–3958.

(66) Giessibl, F. J. The QPlus Sensor, a Powerful Core for the Atomic Force Microscope. *Rev. Sci. Instrum.* **2019**, *90*, 11101.

(67) Rezeq, M.; Pitters, J.; Wolkow, R. Tungsten Nanotip Fabrication by Spatially Controlled Field-Assisted Reaction with Nitrogen. *J. Chem. Phys.* **2006**, *124*, 204716.

(68) Labidi, H.; Koleini, M.; Huff, T.; Salomons, M.; Cloutier, M.; Pitters, J.; Wolkow, R. A. Indications of Chemical Bond Contrast in AFM Images of a Hydrogen-Terminated Silicon Surface. *Nat. Commun.* **2017**, *8*, 14222.

(69) Lyding, J. W.; Shen, T.-C.; Abeln, G. C.; Wang, C.; Tucker, J. R. Nanoscale Patterning and Selective Chemistry of Silicon Surfaces by Ultrahigh-Vacuum Scanning Tunneling Microscopy. *Nanotechnology* **1996**, *7*, 128.

**Supporting Information: Electrostatic Landscape of a H-Silicon Surface Probed by a Moveable Quantum Dot**

**Table of Contents**

-Negative Hydrogen

-Supporting Information Figures 1-7

**Negative Hydrogen**

We show that a lone hydrogen on the H-Si(100)-2×1 surface is negatively charged. A hydrogen atom released by creating a DB sometimes appears to sit at metastable interstitial surface site and induces a lattice distortion of the adjacent dimer pairs as reported in ref. [30] and shown again in Figure S5. Figure S6a shows a STM image of a DB pair with a physisorbed hydrogen in the upper left that landed near the pair after the tip was used to create the top DB. Figure S6b,c are nc-AFM $\Delta f$ maps of the same area while the physisorbed hydrogen is present (dashed white circle), taken at two different heights. In both nc-AFM images, the top DB displays a less negative $\Delta f$ shift than the lower one. Work done on DB pairs in another work [6] showed that a negative charge perturbation biasing a pair of DBs showed the same behavior; the one closer to the negative perturbative charge shows up with a less negative $\Delta f$ shift. This biasing is confirmed by removing the physisorbed hydrogen. Figure S6d shows the STM image after removal, with Figure S6e,f being the nc-AFM images at the same two selected heights now showing the two DBs equal in $\Delta f$ shift. $\Delta f(V)$ spectroscopy taken over the pairs before and after hydrogen removal are shown in Figure S6g. In the presence of the hydrogen, the curves taken above the two DBs (dark and light purple curves) show a shift of the spectra. Two interesting features become apparent. First, the (0/-) charge transition step (marked by the vertical dashed lines), are offset for the two curves due to the slightly different distance from the negative hydrogen. Second, a small secondary dip or step is apparent at -0.36 V for both curves. This step has been reported to be the (+/0) charge transition level of the DB,[24] which due to the perturbation of the hydrogen becomes accessible in the voltage range being probed. After hydrogen removal (orange and pink curves), the (0/-) transition steps return to more negative values and are no longer offset from each other. A common point was identified for both before and after curves (vertical dashed lines). Using these points, analysis of the DB energy shift as a function of distance from the hydrogen was performed and plotted in Figure 3 of the main text as the two purple points.

We note here that unlike the $\mathit{\Delta f(V)}$ spectroscopy presented in the main text, these curves do not have a clean step for the charge transition. To obtain a measurable signal the spectroscopy was taken at a height of $z_{rel}$ = -350 pm, which can distort the shape of the curves due to non-linear interactions with the surface (See Supporting Information, Figure S7). Additionally, it is known that hydrogens from a desorption event sit on a metastable surface site and can easily be moved with the field of a scanning tip.[30] The small tip-sample separation and the mechanical instability of the loosely-bound hydrogen atom as the bias is swept for the $\mathit{\Delta f(V)}$ spectroscopy account for the noisy appearance of the curves. A set of $\mathit{\Delta f(V)}$ curves taken at a larger tip-sample separation over the two DBs without the hydrogen present (blue curves) show a more typical charge transition step.

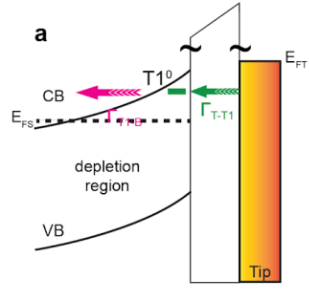

$\Gamma_{T1-B} < \Gamma_{T-T1}$ :T1=Neutral
V= Above CB Edge

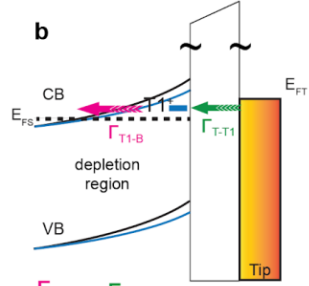

$\Gamma_{T1-B} > \Gamma_{T-T1}$ :T1=Positive
V= Slightly Positive

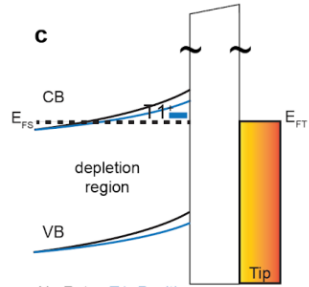

No Rates:T1=Positive
V= 0 V
Upward Band Bending due to CPD

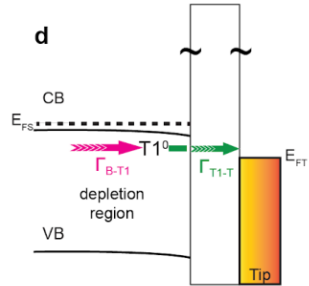

$\Gamma_{B-T1} > \Gamma_{T1-T}$ :T1=Neutral
V= CPD

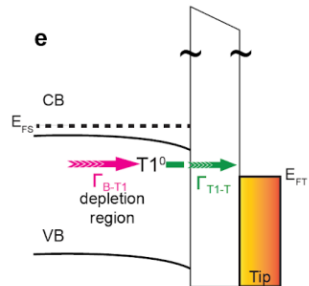

$\Gamma_{B-T1} > \Gamma_{T1-T}$ :T1=Neutral
V= CPD to ~ -1.2 V

**Figure S1: Arsenic Dopant Energy Level Diagrams. (a-e)** Qualitative energy level diagrams of the arsenic dopant as it is swept through the voltage ranges listed underneath each panel. The dopant changes charge state depending on the tip-induced band bending conditions which tune the tunneling rates between tip and T1 ($\Gamma_{T\text{-}T1}$) and bulk and T1 ($\Gamma_{B\text{-}T1}$).

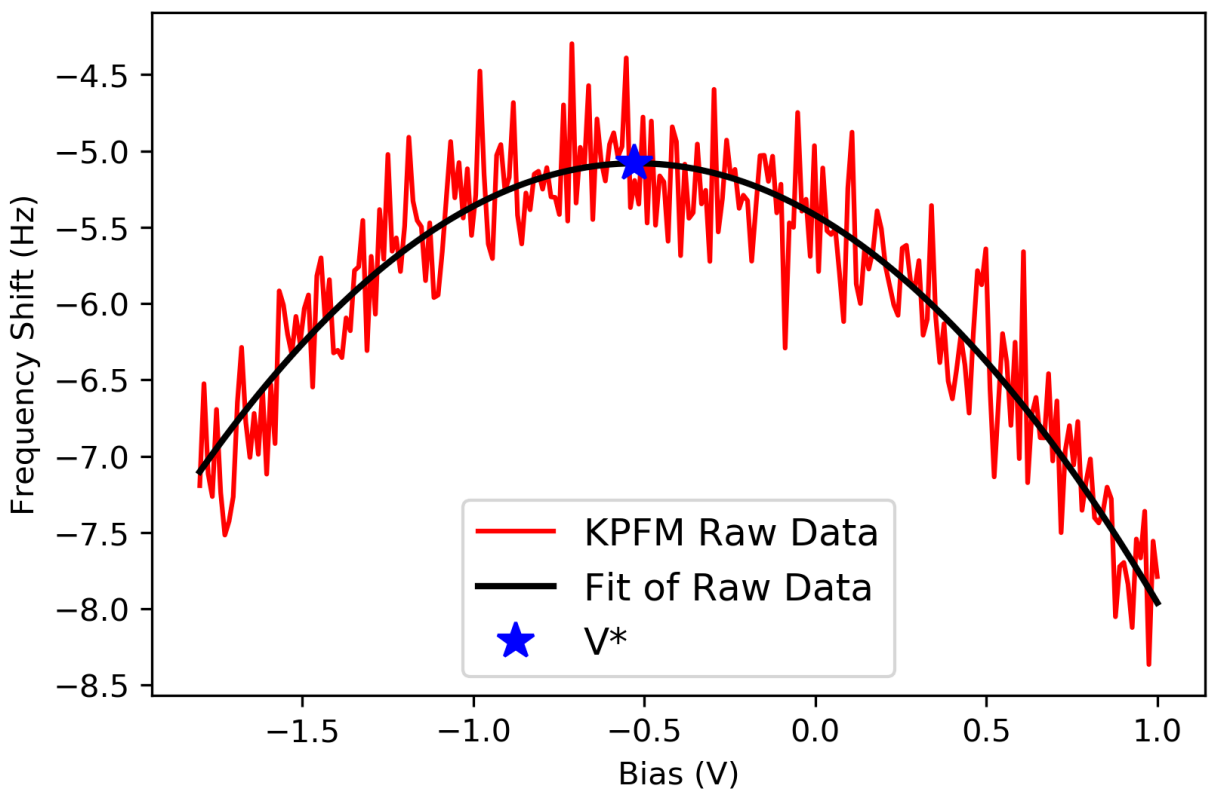

**Figure S2: Fitting a KPFM Curve.** Raw KPFM data from Figure 1n fit with a parabolic function ($z_{rel} = 0.0$ pm, $V_{range} = -1.8$ to $1.0$ V, and Osc. Amp = 50 pm). The maximum of the fit (-0.53 V) is marked with the blue star. Every point in a KPFM grid has this performed. The extracted maxima or V*'s vary slightly for every point, generating a map of the electrostatic potential (*cf.* Figure 1n).

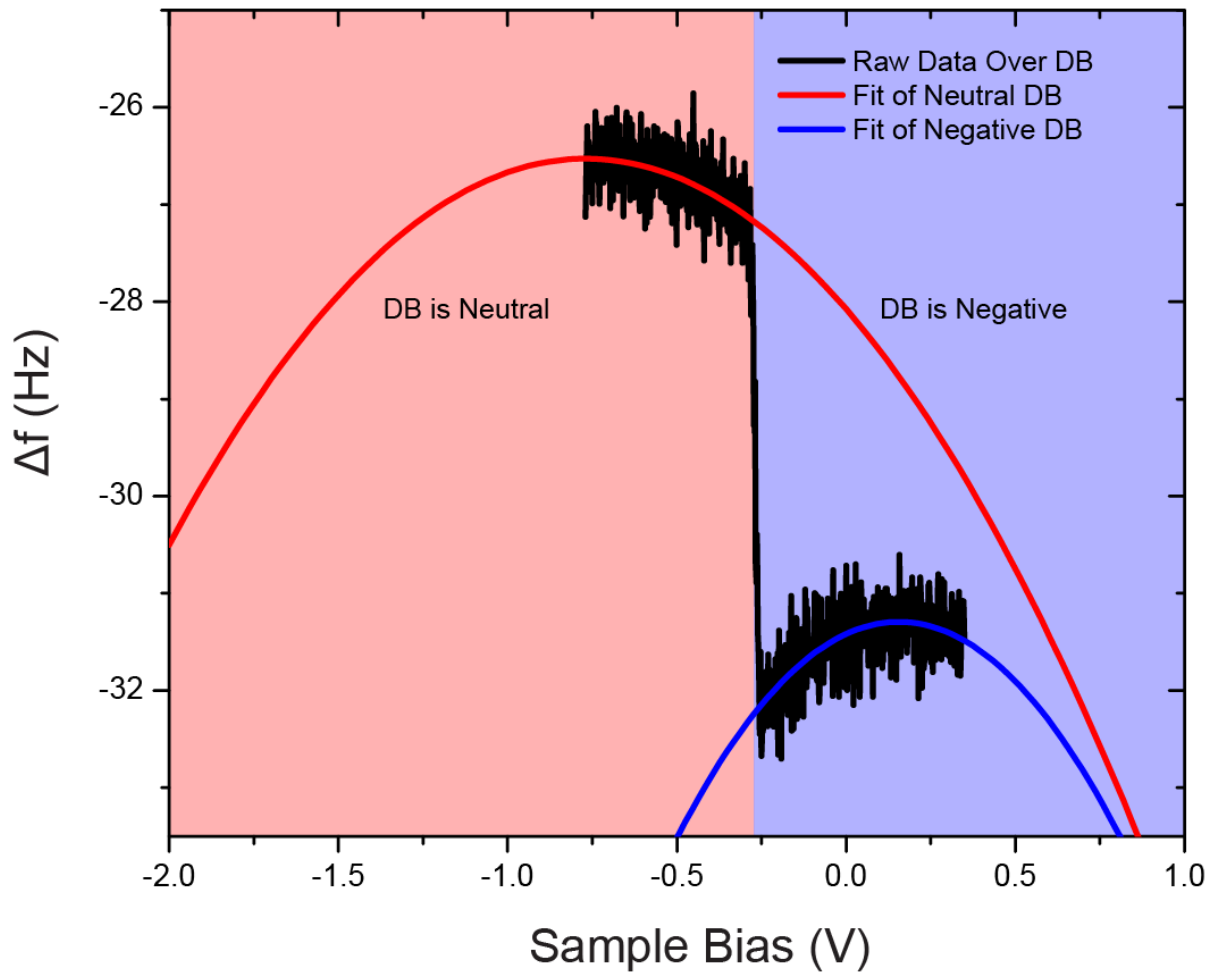

**Figure S3: Fitting a DB KPFM Curve.** Raw KPFM data (black curve) taken over a DB ($z_{rel}$ = -300 pm). For voltages in the red shaded area, the DB is neutral. For voltages in the blue shaded area, the DB is negative. The transition from neutral to negative, or the (0/-) charge transition step is seen at V = -0.25 V. The parabolas were fit to both the neutral (red) and positive (blue) charge state of the DB. The LCPD is extracted from the maximum value for each parabolic fit.

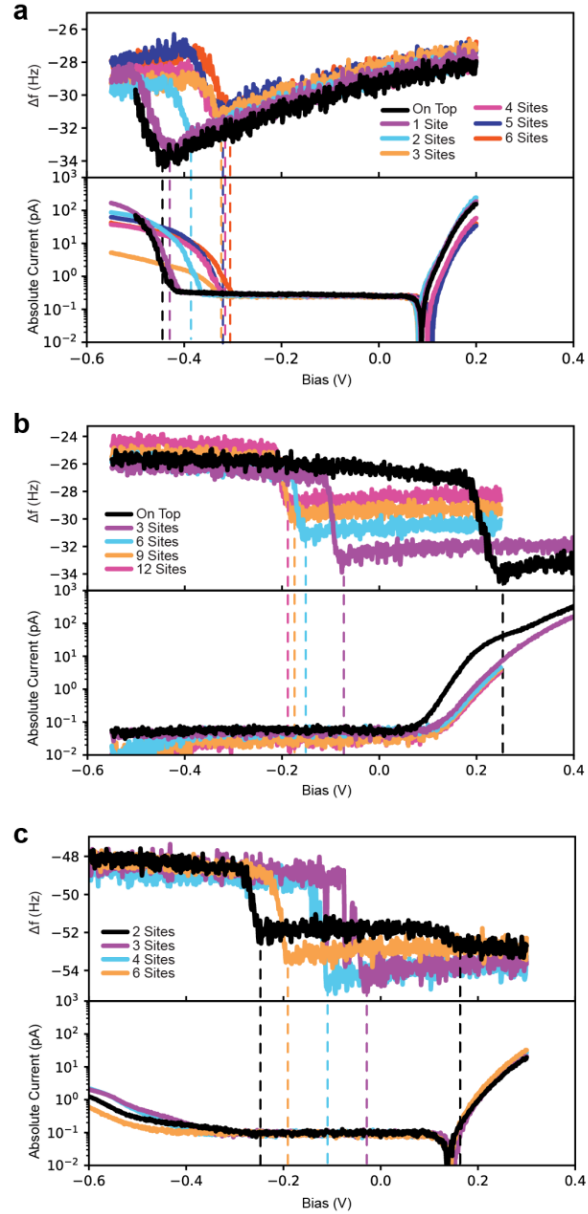

**Figure S4:** *I(V)* **and Extracted Shifts for *Δf(V)*Spectroscopy.** (**a**) The *Δf(V)* spectroscopy from main text Figure 2l for the DB-T1 case is reproduced in the top panel, with the simultaneously obtained *I(V)* spectrum at the bottom. As the DB probe is moved farther away, shifts also occur in the onset of the valence and conduction bands. (**b**) *Δf(V)* spectroscopy from main text Figure 2s for the DB-T2 case, along with its *I(V)* spectroscopy. (**c**) *Δf(V)* spectroscopy from main text Figure 2x for the DB-DB case and its *I(V)* spectroscopy. For all three cases their charge transition step onsets have been marked with a color coded dashed line. These onsets were used as the common point of reference for extracting the electrostatic energy shifts plotted in Figure 3.

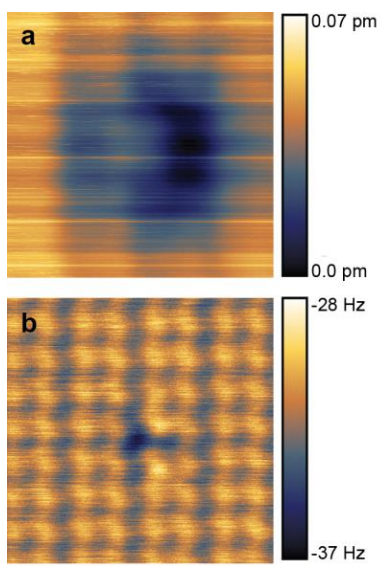

**Figure S5: Lattice Distortion from a Hydrogen Atom. (a)** Constant current empty states (V = 1.3 V and I = 50 pA) STM image of a lone Hydrogen atom on the H-Si surface. **(b)** Constant height AFM image of the same area showing the Hydrogen sits at an apparent surface hollow, distorting the upper hydrogen layer ($z_{rel}$ = -360 pm, V = 0 V, and Osc. Amp = 50 pm). Both images are 3×3 nm$^2$.

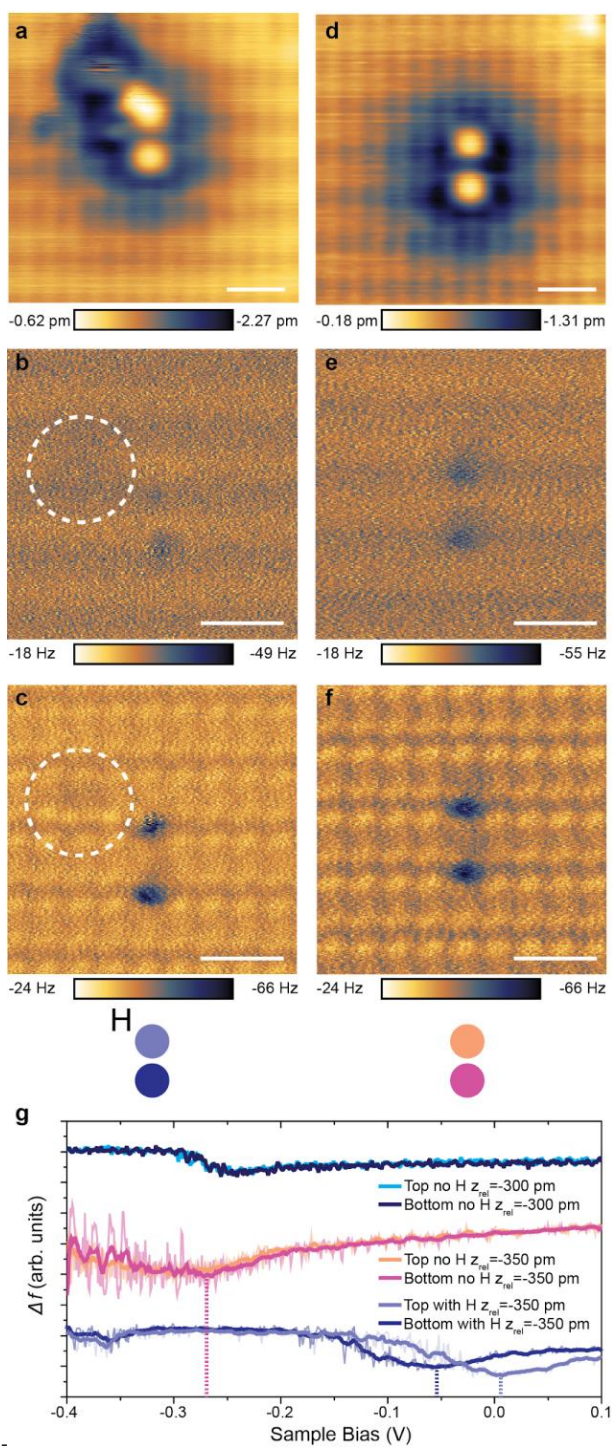

**Figure S6: Negative Hydrogen Atom Perturbing a Pair.** (**a**) Constant current empty states (V = 1.3 V and I = 50 pA) STM image of a DB pair with a physisorbed lone hydrogen visible in the upper left. Constant height AFM images were taken at (**b**) $z_{rel}$ = -300 pm and (**c**) $z_{rel}$ = -350 pm, of the same perturbed pair showing the upper DB lighter in contrast than the lower (V = 0 V, and Osc. Amp = 50 pm). The location of the hydrogen atom is marked by the dashed white circles. (**d**) Constant current empty states (V = 1.3 V and I = 50 pA) STM image

of the pair after removal of the hydrogen. The constant height AFM images were repeated at the same heights of (e) $z_{rel} = -300$ pm and (f) $z_{rel} = -350$ pm (V = 0 V, and Osc. Amp = 50 pm). (g) *Δf(V)* spectroscopy taken on top of the DBs both with the hydrogen present and without. Filtered curves are overlaid on the raw data. Color coded models are provided above the panel. ($z_{rel} = -350$ pm and Osc. Amp = 50 pm). Vertical color coded dashed lines mark a common point of reference. All scale bars in images are 1 nm.

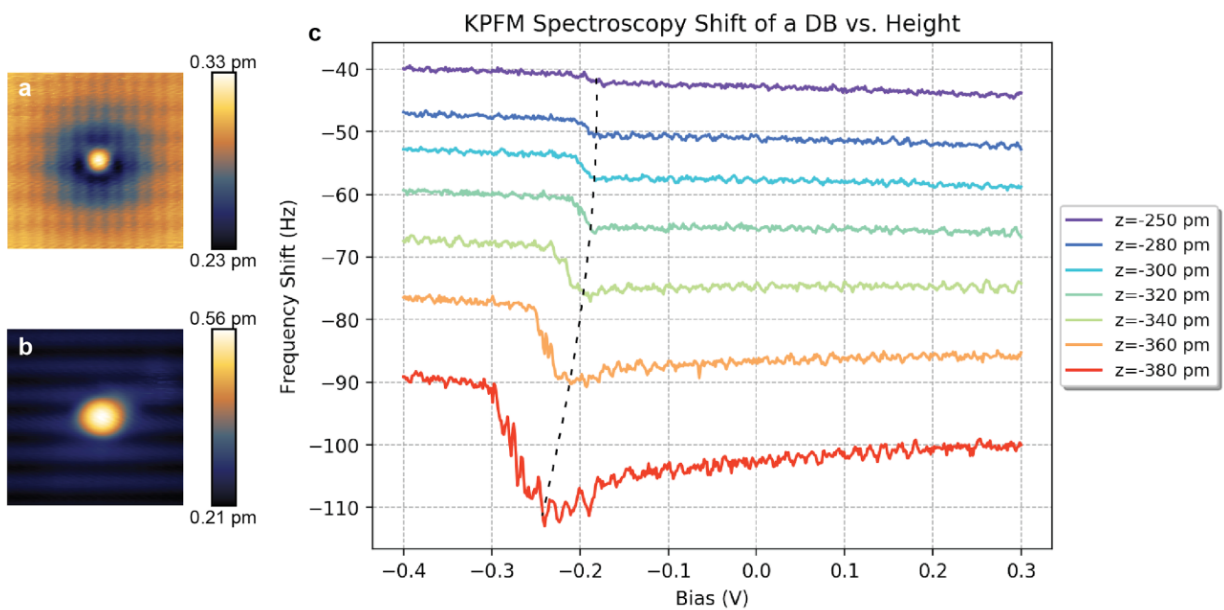

**Figure S7: KPFM Spectroscopy Shift with Height.** Constant current (a) empty states (V = 1.3 V and I = 50 pA) and (b) filled states (V = -1.8 V and I = 50 pA) STM images of a lone DB. (c) KPFM spectroscopy performed over the same DB at different heights ($V_{range}$ = +0.3 to -0.4 V and Osc. Amp = 50 pm), showing a shift of the neutral to negative charge transition level (0/-) as a function of height. Tip-induced band bending shifts the energetic position of the charge transition to more negative values with decreasing tip-sample separation. A black dashed line is provided as a guide to the eye.